\documentstyle[preprint,aps]{revtex}
\begin{document}

\tightenlines
\draft
\preprint{SNUTP 99-015}
\title{Localization instability  in the rotating D-branes}

\author{Rong-Gen Cai and Dahl Park}
\address{Center for Theoretical Physics, Seoul National
       University, Seoul 151-742, Korea}
\author{Kwang-Sup Soh}
\address{Department of Physics Education, Seoul National University,
   Seoul 151-742, Korea}
\maketitle

%%================================================

\begin{abstract}
In the microcanonical ensemble for string theory on $ AdS_m \times S^n$, 
there is a phase transition between a black hole solution extended over the
$S^n$ and a solution localized on the $S^n$ if the $AdS_m$ has the topology 
$R^2 \times S^{m-2}$. The phase transition will not appear if the $AdS_m$ 
has the topology $R^2 \times T^{m-2}$, that is, when the $AdS_m\times S^n$ 
geometry  is regarded as arising from the near-horizon limit of a black
$m-2$ brane. In this paper, we argue that when the black branes are rotating,
the localization phase transition  will occur  between some  rotating 
branes and  corresponding  Kerr black holes when the angular momentum 
reaches its critical value.

\end{abstract}

%%===================================================

\newpage

The Maldacena's  conjecture \cite{Mald} asserts that string/M theory 
on anti-de Sitter (AdS) space (times a compact space) is dual to  certain 
large $N$ conformal field theory living on the boundary of the AdS
(for a recent review see \cite{Review}).
  Thus some questions
concerning large $N$ gauge theories may be answered via supergravity. 
Indeed, one can use this duality to calculate the correlation
functions of gauge theory \cite{Gubser,Witten1}, and to study 
thermodynamics and phase structure of the strong coupling gauge theory 
\cite{Itzhaki,Witten2,Li1,Barbon,Abel,Li2,Martinec,Mart,Gao}.

Recently, Banks, Douglas, Horowitz and Martinec \cite{BDHM} argued that 
in the microcanonical ensemble for string theory on $AdS_m\times S^n$, 
there is a phase transition between  a black hole solution extended 
on the $S^n$ and a solution localized on the $S^n$, which  
is associated with the Gregory-Laflamme instability 
(localization instability) of black strings and $p$-branes \cite{Greg}.
This phase transition arises as follows. 
 It was  shown by Hawking and Page 
\cite{Hawking}, that a large mass Schwarzschild black hole in AdS space 
has positive heat capacity, while a small mass black hole has negative
heat capacity. This implies that in the canonical ensemble the small 
mass black hole is unstable. However, in the microcanonical ensemble
this instability is absent. Instead the system enters into a new  phase, 
in which the black hole localizes on the sphere $S^n$, because the 
Schwarzschild black hole in $m+n$ dimensions  has larger entropy than 
the equal mass Schwarzschild-AdS black hole. Thus,  at higher energies,
 the typical state is a Schwarzschild-AdS black hole with a constant radius
 sphere $S^n$; at lower energies, where the horizon radius is smaller than 
the cosmological scale, the typical state becomes a Schwarzschild black hole
in $m+n$ dimensions. That is, the black hole is localized on the 
sphere $S^n$.

More recently, Peet and Ross \cite{Peet} found  that there is
 a subtlety in the above localization phase transition. They showed
 that the existence of the phase transition crucially depends on the 
spatial boundary condition (topology) of the $AdS_m$ space. The above 
transition may happen when the $AdS_m$ has topology $R^2 \times S^{m-2}$. 
If one thinks of the $AdS_m\times S^n$ geometry as arising from the 
near-horizon limit of some $m-2$ brane, 
the $AdS_m$ space has topology $R^2\times T^{m-2}$ or its large volume 
limit $R^2 \times R^{m-2}$. In this case.  this localization phase 
transition will not  occur: at lower energies, the state is still the 
Schwarzschild-AdS black hole. 
 This is associated with  the fact that in contrast to the
usual Schwarzschild-AdS black hole, the so-called topological black holes
in AdS space always have positive heat capacity \cite{cai}. This  
implies that this localization instability does not appear as well for other
higher genus topological manifold $\Sigma_g^{m-2}$  such  that the 
$AdS_m$ has topology $R^2 \times \Sigma_g^{m-2}$.

As is well known, the near-horizon geometries of the D3-brane, D1-D5-brane,
 M2-brane and M5-brane typically have structures  $AdS_5 \times S^5$, 
$AdS_3\times S^3$, $AdS_4\times S^7$ and $AdS_7\times S^4$, and the $AdS_m$ 
has the topology $R^2\times T^{m-2}$. Therefore, the
localization phase transition does not take place on these branes. In  
the present paper, we are going to  consider minor extension of them,
 that is, rotating D3-branes, M5-branes 
and M2-branes. In \cite{Gubser2} Gubser showed that the low energy excitation
of the rotating black D3-branes is thermodynamically stable up to a critical
angular momentum density and a field theory model can correctly predict this
critical limit. In a previous paper, in the canonical ensemble and grand 
canonical ensemble we investigated this critical behavior  in the rotating 
D3-branes, M2-branes and M5-branes and found that the their heat capacity   
will become negative when the angular momentum  exceeds a certain 
critical value \cite{cai2}.  This leads us to suspect that in the 
microcanonical ensemble the above localization  phase transition may occur 
in the rotating branes. Indeed,
as we will show shortly, at the low energy, but large angular momentum 
regime, a typical state should be  a Kerr black hole, which has a larger 
entropy than the corresponding rotating black brane.

%%===================D3-brane====================

Let us begin our discussion with the rotating black D3-brane in the type IIB
supergravity. The rotating black D3-brane solution has been given 
in \cite{Russo,Kraus,Russo1,Youm}. In general, the black D3-brane solution
 may have three angular momentum parameters. For our purpose, however, 
 it is sufficient to  consider  that only a single angular momentum 
parameter does not vanish.  In this case,  The metric is
\begin{eqnarray}
\label{e1}
ds^2 &=& \frac{1}{\sqrt{f}}\left(-h dt^2 +dx_1^2 +dx_2^2+dx_3^2 \right)
      +\sqrt{f}\left [\frac{dr^2}{\tilde{h}}-\frac{4ml\cosh \alpha}{r^4
      \triangle f}\sin^2\theta dtd\phi
      \right. \nonumber \\
     &+& \left. r^2 (\triangle d\theta^2 +\tilde{\triangle}\sin^2\theta
       d\phi^2 +\cos^2\theta d\Omega_3^2\right],
\end{eqnarray}
where
\begin{eqnarray}
&& f=1+\frac{2m \sinh^2\alpha}{r^4\triangle}, \nonumber \\
&& \triangle =1+\frac{l^2\cos^2\theta}{r^2}, \nonumber\\
&& \tilde{\triangle}=1+\frac{l^2}{r^2} +\frac{2m l^2\sin^2\theta}
         {r^6\triangle f}, \nonumber \\
&& h=1-\frac{2m}{r^4 \triangle}, \nonumber \\
&& \tilde{h}=\frac{1}{\triangle}\left (1 +\frac{l^2}{r^2}
-\frac{2m}{r^4}\right). \nonumber
\end{eqnarray}
The extremal limit of the solution  is approached by letting 
$m\rightarrow 0$.  An interesting point of the solution is  
that the extremal limit of the solution (\ref{e1})
can be interpreted as some superposition of $N$ static D3-branes, rather
than that of $N$ coinciding rotating D3-branes \cite{Kraus,Sfetsos}.
The near-horizon limit of static D3-branes has the geometry $AdS_5 
\times S^5$ with same radii $b_3^2=l_s^2 (4\pi g_s N)^{1/2}$ for the
$AdS_5$ and $S^5$, where $l_s$ and $g_s$ are string length and coupling
constant, respectively. For the rotating D3-brane solution (\ref{e1}),
the geometry $AdS_5\times S^5$ is twisted because of the rotation.
Near the extremal limit, some  thermodynamic quantities  of the black 
three-branes was given in \cite{cai2}:
\begin{eqnarray}
\label{e2}
&& E=3 \pi ^3 \kappa^{-2} m L^3,  \nonumber \\
&& J= \pi ^{7/4}\kappa^{-3/2}N^{1/2}m^{1/2}lL^3, \nonumber  \\
&& \Omega =\pi^{5/4}\kappa^{-1/2}N^{-1/2}m^{-1/2}l r_+^2, \nonumber  \\
&& T=2^{-1}\pi^{1/4}\kappa^{-1/2}N^{-1/2}m^{-1/2}(2r^3_+ +l^2r_+),
 \nonumber \\
&&  S=2 \pi^{11/4}\kappa^{-3/2}N^{1/2}m^{1/2}r_+L^3.
\end{eqnarray}
Here $E$ denotes the energy above the extremality which equals the 
ADM mass of the black three-brane minus the mass of the 
corresponding extremal one; $L^3$ is the spatial volume of the world 
volume; $2\kappa^2= (2\pi )^7 g_s^2 l_s^8 $; $N$ is the number of the 
coincident D3-branes; and $J$, $\Omega$, $T$ and $S$
represent the angular momentum, angular velocity, Hawking temperature,
and the entropy, respectively. Finally, $r_+$ is the horizon radius of the
black three-branes, which is 
\begin{equation}
\label{e7}
r_+^2=\frac{1}{2}\left (\sqrt{l^4+8m}-l^2\right).
\end{equation}
In \cite{cai2} we showed that in the grand canonical ensemble the heat 
capacity becomes negative when $l^4/m  >8/3$, which indicates there is a 
critical point at $l^4/m=8/3$. In the canonical ensemble there is also
 a similar critical point. Therefore  the black three-brane is unstable 
when  the angular momentum exceeds a certain critical value. According to
 the Maldacena's conjecture, this also implies that there is a phase 
transition in the ${\cal N}=4$ large $N$ super Yang-Mills theory in four 
dimensions at finite temperature \cite{Gubser2,cai2}.
 In the microcanonical ensemble, where the energy $E$ and the angular 
momentum $J$ are fixed, however, similar to the case for the small mass 
Schwarzschild-AdS black hole, the above instability 
is  absent. Instead a localization instability mentioned above will appear.
That is, at low energy and large angular momentums, that is, when the horizon
radius is smaller than the cosmological scale, 
the system enters into another  entropically  favored state, 
which,  according to the no-hair theorem of black holes, 
should be described by  a Kerr black hole. In addition, the D3-branes 
distributed on a planar disc \cite{Kraus} might  be an appropriate source 
for the Kerr solution.

%%====================Kerr black hole====================== 

A D-dimensional ($D\ge 4$) Kerr black hole solution with an angular momentum
parameter was given in \cite{Myers}. In ``Boyer-Lindquist'' coordinates, its
metric is 
\begin{eqnarray}
ds^2= &-& dt^2 +\sin^2\theta (r^2+a^2)d\phi^2 +\triangle (dt 
          +a \sin^2d\phi^2)^2 \nonumber \\
     &+& \Psi dr^2 + \rho ^2 d\theta^2 +r^2 \cos^2\theta d\Omega^2_{D-4},
\end{eqnarray}
where
\begin{eqnarray}
&& \triangle =\frac{\mu}{r^{D-5}\rho ^2},\ \ \rho^2 =r^2 +a^2 \cos^2\theta, 
          \nonumber \\
&& \Psi =\frac{r^{D-5}\rho ^2}{r^{D-5}(r^2 +a^2)-\mu}. \nonumber
\end{eqnarray}
This solution has the ADM mass
\begin{equation}
M_k=\frac{(D-2)\Omega_{D-2}}{2\kappa ^2}\mu,
\end{equation}
and the angular momentum 
\begin{equation}
J_k= \frac{\Omega_{D-2}}{\kappa ^2}\mu a =\frac{2a}{D-2}M_k.
\end{equation}
Here $\Omega_{D-2}=2 \pi^{(D-1)/2}/\Gamma [(D-1)/2] $ is the area of a unit 
$(D-2)$-sphere. The horizon $r_k$ of the Kerr black hole is determined 
by the equation $g^{rr}=0$, that is,
\begin{equation}
\label{e11}
r^{D-5}_k (r^2_k +a^2 ) -\mu=0.
\end{equation}
And the entropy of the solution is one quarter horizon area,
\begin{equation}
\label{kerrentropy}
S_k =2 \pi (r_k^2 +a^2 )r_k^{D-4} \Omega_{D-2}/\kappa^2.
\end{equation}
Here it seems worth pointing out that for the higher dimensional ($D>5$) Kerr
solutions there is no any restriction on the angular momentum 
in order for the solutions  to have the  black hole horizon.

Now Let us compare the entropies of the rotating black three brane 
(\ref{e2}) and a ten-dimensional  Kerr black hole (\ref{kerrentropy}).
These two entropies can be compared extactly.
 For small mass and large angular momentum, we find
\begin{eqnarray}
\label{e13}
&&   S \sim \frac{b_3^8}{\kappa^2 }\left (\frac{\kappa^2 J}{b_3^8} 
     \right )^{-1} \left (\frac{\kappa ^2 M}{b_3^7}
       \right)^{3/2}, \nonumber \\ 
&&   S_k \sim \frac{b_3^8}{\kappa^2 }\left(\frac{\kappa^2 J_k}{b^8_3 }
        \right )^{-2/5}
     \left (\frac{\kappa^2 M_k}{b_3^7}\right)^{8/5},
\end{eqnarray} 
where following \cite{Peet} we have rescaled the energy of excitations $E$ as 
$M=LE/b_3$. From the above we can see that these two entropies are comparable 
when $\kappa^2 M \sim b_3^7$ and $\kappa^2 J \sim b_3^8$. Note that in this 
comparison we should take $M=M_k$ and $J=J_k$. When $\kappa^2 M \sim b_3^7$,
for large angular momentum $\kappa^2 J >b_3^8$, the Kerr black hole has larger
entropy than the rotating D3-brane.  Therefore, a localization 
phase transition may happen when $\kappa^2 M \sim b^7_3$ 
and $\kappa^2J \sim b_3^8$. In the phase diagram of $J$ versus $M$, the 
critical line of the localization  phase transition is $ \left 
(\frac{\kappa ^2 J}{b_3^8}\right )^6 \left (\frac{\kappa ^2 M}{b_3^7}
\right ) \sim 1$. The phase above the critical line is described by the 
rotating D3-branes, while the phase below the line by the Kerr black hole. 
Of course, it should be pointed out here that there are other phases in this 
phase diagram, which will be discussed at the end of the paper.

%%===================================M5-brane===============

Now we discuss the rotating M5-branes in the eleven-dimensional supergravity.
 Similar to the case of the  rotating D3-brane, we also consider
the case in which an angular momentum parameter does not vanish only.  The
rotating M5-brane solution has been given   in \cite{Cvetic,Csaki}.
The metric  can be written down as
\begin{eqnarray}
\label{2e1}
ds^2_{11} &=& f^{-\frac{1}{3}}(-hdt^2 +dx_1^2 +\cdots +dx_5^2) +
       f^{\frac{2}{3}}\left [\frac{dr^2}{\tilde{h}} + r^2 (\triangle
       d\theta^2 +\tilde{\triangle}\sin^2 \theta d\phi^2
       \right. \nonumber \\
  &+& \left. \cos^2\theta d\Omega^2_2)
    -\frac{4ml \cosh\alpha}{r^3\triangle f}\sin^2\theta dtd\phi \right],
\end{eqnarray}
where
\begin{eqnarray}
&& f=1+\frac{2m \sinh^2\alpha}{r^3\triangle}, \nonumber \\
&& \triangle =1+\frac{l^2\cos^2\theta}{r^2}, \nonumber \\
&& \tilde{\triangle}=1+\frac{l^2}{r^2} +\frac{2m l^2
      \sin^2\theta}{r^5\triangle f}, \nonumber \\
&& h=1-\frac{2m}{r^3\triangle}, \nonumber\\
&& \tilde{h}=\frac{1}{\triangle}\left[1+\frac{l^2}{r^2}-\frac{2m}{r^3}\right].
    \nonumber
\end{eqnarray}
The horizon of the rotating black M5-branes is determined by the equation
\begin{equation}
\label{e16}
r_+^3 +l^2 r_+ -2m=0.
\end{equation}
Near the extremal limit, some  thermodynamic quantities of the black M5-branes
are \cite{cai2}
\begin{eqnarray}
\label{e17}
&& E=20 \pi ^2  3^{-1}   \kappa^{-2} m L^5, \nonumber \\
&& J=2^{7/3} 3^{-1}\pi ^{7/6} \kappa^{-5/3} N^{1/2}m^{1/2}l L^5,
    \nonumber  \\
&& \Omega =2^{2/3}\pi^{5/6}\kappa^{-1/3}N^{-1/2}m^{-1/2}lr_+,
    \nonumber  \\
&& T=2^{-4/3}\pi ^{-1/6}\kappa^{-1/3}N^{-1/2}m^{-1/2}(3r_+^2+l^2),
    \nonumber  \\
&& S=2^{10/3} 3^{-1} \pi ^{13/6}\kappa^{-5/3}N^{1/2}m^{1/2}r_+ L^5,
\end{eqnarray}
where $2\kappa ^2=(2\pi)^8 l_p^9$ and $l_p$ is the eleven-dimensional 
Planck  constant. In the case of static M5-branes, the radius of $AdS_7$
is $b_5=2 l_p (\pi N)^{1/3}$.  Under the large angular momentum 
approximation, the entropy (\ref{e17}) of the rotating M5-branes behaves as
\begin{equation}
S \sim \frac{b_5^9}{\kappa^2} \left (\frac{\kappa^2 J}{b_5^9}
     \right)^{-2}\left (\frac{\kappa^2 M}{b_5^8}\right)^{5/2},
\end{equation}
while a eleven-dimensional, large angular momentum Kerr black hole has 
entropy behavior:
\begin{equation}
\label{entropy11}
S_k \sim \frac{b_5^9}{\kappa^2}\left (\frac{\kappa^2 J_k}{b_5^9}
      \right)^{-1/3} \left (\frac{\kappa ^2 M_k}{b_5^8}\right)^{3/2}.
\end{equation}
When $M=M_k$ and $J=J_k$, obviously, these two entropies are 
comparable as well if $\kappa^2 M \sim b_5^8$ and $\kappa^2 J\sim 
b_5^9$.   When $\kappa^2 M \sim b^8_5$, for larger angular momentum
$\kappa^2 J >b_5^9$, the Kerr black hole has larger entropy than the 
rotating M5-branes; when keeping $\kappa^2 J\sim b^9_5$, and 
 lowering the mass such that $\kappa^2 M< b_5^8$, we can also see 
that the Kerr black hole may have larger entropy than the rotating 
M5-branes. We therefore conclude that the localization phase transition 
may happen in the rotating M5-branes when $\kappa^2 M \sim b_5^8$ and
$\kappa ^2 J \sim b_5^9$.  That is, at low energies, beyond a critical 
angular momentum, the Kerr black hole is the entropically favored phase.
In this case, the critical line is $\left (\frac{\kappa^2 J}{b_5^9}
\right)^{-5/3}\left(\frac{\kappa^2 M}{b_5^8}\right) \sim 1$. Above the line 
is the Kerr black hole phase and the phase below the line is described by
the M5-branes. Once again, there are also other phases in this phase 
diagram.

%%================M2-brane===========================

We now turn to  the rotating black M2-branes.
The rotating black M2-brane metric with a non-vanishing angular momentum
is \cite{Cvetic}
\begin{eqnarray}
\label{3e1}
ds_{11}^2 &=& f^{-2/3} (-hdt^2 +dx^2_1+dx^2_2) + f^{1/3}\left [ \frac{dr^2}
     {\tilde{h}} +r^2 (\triangle d\theta^2 +\tilde{\triangle}
     \sin^2\theta d\phi^2  \right. \nonumber \\
   &+& \left. \cos^2\theta d\Omega^2_5 )- \frac{4ml \cosh\alpha}
     {r^6\triangle f} \sin^2\theta dtd\phi \right],
\end{eqnarray}
where
\begin{eqnarray}
&& f= 1+\frac{2m\sinh^2\alpha}{r^6\triangle}, \nonumber \\
&& \triangle = 1+\frac{l^2\cos^2\theta}{r^2}, \nonumber \\
&& \tilde{\triangle}= 1+\frac{l^2}{r^2} +\frac{2ml^2\sin^2\theta}
       {r^8 \triangle f}, \nonumber \\
&& h=1-\frac{2m}{r^6\triangle}, \nonumber \\
&& \tilde{h}= \frac{1}{\triangle}\left (1+\frac{l^2}{r^2}-
        \frac{2m}{r^6}\right). \nonumber
\end{eqnarray}
This solution has horizon $r_+$ given by
\begin{equation}
\label{e26}
r_+^6 +r_+^4l^2 -2m =0.
\end{equation}
In this case,  near the extremal limit, we have the following 
thermodynamic behavior of the rotating black M2-brane \cite{cai2}:
\begin{eqnarray}
\label{e27}
&& E=3^{-1} 4\pi ^4  \kappa^{-2} m L^2, \nonumber \\
&& J= 2^{2/3} 3^{-1}\pi ^{7/3} \kappa^{-4/3}N^{1/2}m^{1/2}lL^2, 
    \nonumber \\
&& \Omega = 2^{-2/3}\pi ^{5/3}\kappa^{-2/3}N^{-1/2}m^{-1/2}lr_+^4, 
   \nonumber \\
&& T= 2^{-5/3}\pi ^{2/3}\kappa^{-2/3}N^{-1/2}m^{-1/2}r_+^3(3r_+^2 +2l^2),
    \nonumber  \\
&& S = 2^{5/3}3^{-1}\pi^{10/3}\kappa^{-4/3}N^{1/2}m^{1/2}r_+L^2.
\end{eqnarray}
 Under the low energy and large  angular momentum approximation, 
from (\ref{e27}) we have
\begin{equation}
S \sim \frac{b_2^9}{\kappa^2} \left (\frac{\kappa^2 J}{b_2^9}
     \right)^{-1/2} \left(\frac{\kappa^2 M}{b_2^8}\right), 
\end{equation}
where $b_2=l_p(2^5\pi^2 N)^{1/6}/2$ is the radius of near-horizon
geometry $AdS_4$ of the static M2-branes.  Comparing with the entropy 
(\ref{entropy11}) of the eleven-dimensional Kerr black hole, in this case, 
when $\kappa^2 M\sim b_2^8$ and $\kappa^2 J \sim b_2^9$, we find that these
two entropies are also comparable to one another. When $\kappa^2 M
\sim b_2^8$, the entropy of the  Kerr black hole is larger than 
that of the rotating M2-branes if $\kappa^2 J >b_2^9$. Hence we argue 
that the localization  phase transition may also take place for the rotating
M2-branes at $\kappa^2 M \sim b_2^8$ and $\kappa^2 J\sim b_2^9$.
The critical line is $\left (\frac{\kappa^2 J}{b_2^9}\right) 
\left(\frac{\kappa^2 M}{b_2^8}\right)^3 \sim 1$. In this case, above the 
line is the M2-brane phase and the Kerr black hole describes the phase
 below the line.

%%==============================summary ================

In summary, we have discussed  the microcanonical phases  
on the rotating D3-, M5-, and M2-branes, paying attention
to the so-called localization instability. At low energies,
 beyond a certain critical angular momentum, we have argued that
 the system enters into another entropically favored phase: the phase
 is described by a high dimensional Kerr black hole.
That is, there is a localization phase transition from some rotating 
black branes to corresponding Kerr black holes when the angular momentum
 reaches its critical value.  Note that the localization transition 
does not occur for the static D-branes and M-branes \cite{Peet}. Therefore,
 this is completely an  effect of angular momentum.
Here we should mention that for the rotating 
D1-D5-branes, Martinec and Sahakian \cite{Mart} have already 
argued the existence of  such kind of localization phase transition 
from $AdS_3 \times S^3$ to a six-dimensional (6D) black hole, which is 6D
Schwarzschild black  hole spinning on the $S^3$. Angular momentum is
 introduced in this phase by spinning up the hole along an orbit on the 
equator of $S^3$ with momentum $P \sim J/b$, where $b$ denotes the 
cosmological radius of the $AdS_3$. It is interesting to notice the 
difference between the rotating D1-D5-brane system and the rotating branes
considered in this paper. The near-horizon limit of the 
rotating D1-D5-branes is the rotating BTZ black hole. The latter is always 
thermodynamically stable and has a restriction on the angular momentum with 
 $J\le Mb$, beyond which the BTZ solution does not represent a black hole.   
From \cite{Gubser2,cai2}, we know that the rotating D3-, M5- and M2-branes 
are thermodynamically unstable for larger angular momentum and there is no
any restriction on the angular momentum in order for the branes to have a 
horizon. Thus a non-rotating black hole with a velocity on the $S^n$,
 which has also a restriction on the angular momentum \cite{Mart}, might be 
inappropriate to describe the new phase for these rotating branes. Similar
 to the Schwarzschild  black hole for the low energy phase of string on the
 $AdS_m \times S^n$, the higher dimensional Kerr black hole might be
 reasonable, as discussed in the above.

  Of course, there exist also other phases in the 
microcanonical ensemble of strings on the rotating D-branes. For instance,
as shown in \cite{BDHM}, we expect that when the horizon size of the
 Kerr black hole falls into the regime of the string scale, the system 
goes into a Hagedorn, or long string phase, where the fundamental
 string dominates the entropy, 
this is associated with the correspondence principle of Horowitz and
Polchinski \cite{Horowitz}.  At the lower energies and angular momentums, 
another new phase may occur, where a gas of gravitons propagating in 
a rotating AdS space dominate the entropy. It would be interesting to make
a detailed analysis of the microcanonical phases of string theory on
the D-branes. In addition, in this paper we have analyzed only the 
rotating branes with a single angular momentum parameter. We expect 
that similar localization instability exists in  rotating branes 
with multiple angular momentum parameters, because their thermodynamics
is similar \cite{CG}.

%%==============================================
{\bf  Acknowledgments:} R.G.Cai would like to thank Prof. E. Martinec
 for valuable correspondence and Prof. M. Yu for helpful discussion. We are
 grateful to the referee for very stimulating comments  which help  deepen
 our understanding of the localization phase transition in the microcanonical
 ensemble of string theory.  This work was supported by the 
KOSEF through the CTP at Seoul National University.

\end{document}